\RequirePackage[compress]{cite} 
\documentclass[final,twocolumn,3p,nofootinbib,hyperref]{elsarticle}

\makeatletter
\newcommand*{\rom}[1]{\expandafter\@slowromancap\romannumeral #1@}
\makeatother

\usepackage{lineno,hyperref}
\usepackage{multirow}
\usepackage[utf8]{inputenc}

\usepackage{amssymb}
\usepackage{amsmath}
\usepackage{graphicx,color}

\usepackage{cite}

\usepackage{xspace}	

\newcommand{\roots} {\mbox{$\sqrt{\textit{s}_{NN}}$}\xspace}
\newcommand{\GeVc} {\mbox{GeV/$\textit{c}$}\xspace}

\def  \etas       {\mbox{$\eta / \textit{s}$  }\xspace}

\def \sc23  {\mbox{$\mathrm{SC}(2,3)$   }\xspace}
\def \sc24  {\mbox{$\mathrm{SC}(2,4)$   }\xspace}

\def \nsc23 {\mbox{$\mathrm{NSC}(2,3)$}\xspace}
\def \nsc24 {\mbox{$\mathrm{NSC}(2,4)$}\xspace}

\journal{Phys. Lett. B}
\begin{document}
\title{Model investigations of the correlation between the mean transverse momentum and anisotropic flow in shape-engineered events}
\medskip
\author{Niseem~Magdy$^{1}$
 and
 Roy~A.~Lacey$^{2}$
 } 
\address{$^{1}$Department of Physics, University of Illinois at Chicago, Chicago, Illinois 60607, USA \fnref{xxx}}
\address{$^{2}$Department of Chemistry, State University of New York, Stony Brook, New York 11794, USA}

\fntext[xxx]{niseemm@gmail.com}

\begin{abstract}
The correlation between the event mean-transverse momentum $[p_{\mathrm{T}}]$, and the anisotropic flow magnitude $v_n$, $\rho(v^{2}_{n},[p_{T}])$,  has been argued to be sensitive to the initial conditions in heavy-ion collisions. We use simulated events generated with the AMPT and EPOS models for Au+Au at $\sqrt{\textit{s}_{NN}}$ = 200 GeV, to investigate the model dependence and the response and sensitivity of the $\rho(v^{2}_{2},[p_{T}])$ correlator to collision-system size and shape, and the viscosity of the matter produced in the collisions. We find good qualitative agreement between the correlators for the string melting version of the AMPT model and the EPOS model.   The model investigations for shape-engineered events as well as events with different viscosity ($\eta/s$), indicate that  $\rho(v^{2}_{2},[p_{T}])$ is sensitive to the initial-state geometry of the collision system but is insensitive to sizable changes in $\eta/s$ for the medium produced in the collisions. These findings suggest that precise differential measurements of $\rho(v^{2}_{2},[p_{T}])$ as a function of system size, shape, and beam-energy could provide more stringent constraints to discern between initial-state models and hence, more reliable extractions of $\eta/s$.
\end{abstract}
\maketitle
 A central objective of the current heavy-ion programs at the Large Hadron Collider (LHC) and the Relativistic Heavy-Ion Collider (RHIC) is to understand the transport properties of the quark-gluon plasma (QGP)~\cite{Shuryak:1978ij,Shuryak:1980tp,Muller:2012zq} formed in high-energy heavy-ion collisions. In recent years, particular attention has been given to precision extraction of the specific shear viscosity of the QGP - the ratio of shear viscosity $\eta$, to the entropy density $s$, ($\eta/s$). The specific shear viscosity encodes the ability of the QGP to transport momentum. 
In general observables that characterize the azimuthal anisotropy of particles emitted in the transverse plane, are among the key measurements that have been used to constrain the viscous hydrodynamic response to the initial spatial distribution in energy density, produced in the early stages of the collision~\cite{Danielewicz:1998vz,Ackermann:2000tr,Adcox:2002ms,Heinz:2001xi,Huovinen:2001cy,Hirano:2002ds,Shuryak:2003xe,Hirano:2005xf,Romatschke:2007mq,Luzum:2008cw,Bozek:2009dw,Song:2010mg,Qian:2016fpi,Schenke:2011tv,Teaney:2012ke,Gardim:2012yp,Lacey:2013eia,Magdy:2020gxf}.

These studies indicated that a significant uncertainty in the  $\eta/s$ extractions stems from the uncertainty in the estimates for the initial-state eccentricities employed in the model calculations. Subsequently, several works have sought to construct and investigate new observables insensitive to $\eta/s$ and more sensitive to initial-state effects leading to new constraints for the initial-state models~\cite{Bozek:2016yoj,Giacalone:2017uqx}.
 
 

One such observable, that leverages the correlation between the n$^{\rm th}$-order flow harmonics $v_{n}$, and the average transverse momentum of particles in an event $[p_{T}]$, is the correlation coefficient $\rho(v^{2}_{n},[p_{T}])$~\cite{Giacalone:2020byk,Bozek:2016yoj,Bozek:2020drh,Schenke:2020uqq,Giacalone:2020dln,Lim:2021auv,ATLAS:2021kty};
\begin{eqnarray}
    \rho(v^{2}_{n},[p_{T}]) = \frac{{\rm cov}(v_{n}^{2},[p_T])}{\sqrt{{\rm Var}(v_{n}^{2})} \sqrt{{\rm Var}([p_T])}}\label{eq:1}.
\end{eqnarray}
 Here, $v_{n}$ is eccentricity-driven and the $[p_{T}]$ is related to the transverse size of the overlap region, so events that have similar energy-density but smaller initial-state transverse size should have a larger radial expansion and consequently larger mean transverse momentum~\cite{Bozek:2012fw}. It has also been proposed that the $\rho(v^{2}_{n},[p_{T}])$ correlator is sensitive to the correlations between the initial size and the initial-state deformation of colliding nuclei~\cite{Giacalone:2019pca,Giacalone:2020awm,ATLAS:2021kty}.

Initial measurements of the $\rho(v^{2}_{n},[p_{T}])$ correlator have been reported for p+Pb and Pb+Pbcollisions at  \roots = 5.02~TeV  by the ATLAS Collaboration~\cite{Aad:2019fgl}. 
In Pb+Pb collisions the leading order experimental trend for $\rho(v^{2}_{2},[p_{T}])$ reflected negative values in peripheral or very low multiplicity events, but increased with centrality to positive values for mid to central collisions.
These measurements provided important insights for initial-state models at LHC energies.
The $\rho(v^{2}_{2},[p_{T}])$  correlator has also been studied in hydrodynamic and transport models~\cite{Bozek:2021zim,Bozek:2020drh,Giacalone:2020dln,Lim:2021auv}. Nonetheless, further detailed studies of $\rho(v^{2}_{n},[p_{T}])$ are required to optimize its utility to discern between different inital-state models.

 In this work, we use detailed simulations with both the AMPT~\cite{Lin:2004en}, and EPOS~\cite{Drescher:2000ha,Werner:2010aa,Werner:2013tya} models for Au+Au at \roots = 200 GeV, to study the $\rho(v^{2}_{2},[p_{T}])$ correlators model dependence and its response and sensitivity to the magnitude of  $\eta/s$  and initial-state geometry. We exploit the technique of event-shape engineering to obtain a more detailed influence of the initial-state geometry. Our study emphasizes investigations for Au+Au collisions at \roots = 200 GeV in anticipation of the need for model predictions to compare to upcoming experimental measurements.

This study is performed with simulated events for  Au+Au collisions at \roots = 200~GeV, obtained with the AMPT~\cite{Lin:2004en}, and EPOS~\cite{Drescher:2000ha,Werner:2010aa,Werner:2013tya} models. Computations were performed  for charged hadrons in the transverse momentum  range $0.2 < p_T < 2.0$ GeV/$c$ and the pseudorapidity acceptance $|\eta|$ $<$ $1.0$. The latter choice mimics the acceptance of the STAR experiment at RHIC. 

\begin{itemize}
\item{AMPT Model}: The AMPT  model(version 2.26t9b)~\cite{Lin:2004en} has been widely used to study relativistic heavy-ion collisions at LHC and RHIC energies~\cite{Lin:2004en,Ma:2016fve,Haque:2019vgi,Bhaduri:2010wi,Nasim:2010hw,Xu:2010du,Magdy:2020bhd,Guo:2019joy,Magdy:2021sip,Magdy:2020fma}. 
 In the current study, simulations were made with the string melting option both on and off. In such a situation when the string melting mechanism is on, hadrons created from the HIJING model are converted to their valence quarks and anti-quarks, and their evolution in time and space is then shaped by the ZPC parton cascade model~\cite{Zhang:1997ej}.
The key elements of AMPT include (i) HIJING model~\cite{Wang:1991hta,Gyulassy:1994ew} initial parton-production stage  , (ii) a parton scattering stage, (iii) hadronization through coalescence  followed by (iv)  a hadronic interaction stage~\cite{Li:1995pra}.  
In stage (ii) the used parton scattering  cross-sections are evaluated according to;
\begin{eqnarray} \label{eq:21}
\sigma_{pp} = \dfrac{9 \pi \alpha^{2}_{s}}{2 \mu^{2}},
\end{eqnarray}
where $\alpha_{s}$ is the QCD coupling constant and $\mu$ is the screening mass in the partonic matter. 
They principally establish the expansion dynamics of the A+A collision systems~\cite{Zhang:1997ej}; 
Within the AMPT model framework, the $\eta/s$ value can be adjusted via an appropriate choice of $\mu$ and/or $\alpha_s$ for a particular initial temperature $T_{i}$~\cite{Xu:2011fi}.
\begin{eqnarray} \label{eq:22}
 \dfrac{\eta}{s} = \dfrac{3 \pi}{40 \alpha^{2}_{s}}  \dfrac{1}{ \left(  9 +  \dfrac{\mu^2}{T^2} \right)  \ln\left(\dfrac{18 + \mu^2/T^2}{ \mu^2/T^2 } \right) - 18} \nonumber,\\
\end{eqnarray}
In the current work, Au+Au collisions at $\sqrt{s_{\rm NN}}=$ 200~GeV, were simulated with model version ampt-v2.26t9b 
 for a fixed value  $\alpha_{s}$ = 0.47, but the shear viscosity $\eta/s$ is varied over the range 0.1--0.3 by adjusting  $\mu$ from  2.26 -- 4.2~$fm^{-1}$ for a temperature  $T_{i}$ = 378~MeV~\cite{Xu:2011fi}. The three AMPT sets which will be presented in this work are summarized in Tab.~\ref{tab:1}.
\begin{table}[h!]
\begin{center}
 \begin{tabular}{|c|c|c|c|}
 \hline 
 AMPT-set  &        ~ $\eta/s$~          &                   String Melting  Mechanism       \\
  \hline
  Set-1        &         ~  0.1~                &                     OFF                        \\
 \hline
  Set-2       &          ~  0.1~                &                     ON                         \\
 \hline 
  Set-3        &         ~  0.3 ~               &                     ON                         \\
 \hline 
\end{tabular} 
\caption{The summary of the three AMPT sets which will be presented in this work.}
\label{tab:1}
\end{center}
\end{table}

\item{EPOS Model}: 
The EPOS model~\cite{Drescher:2000ha,Werner:2010aa,Werner:2013tya} is based on a 3+1D viscous hydrodynamical description of A+A collisions. The initial state conditions are defined in terms of flux tubes estimated via Gribov-Regge multiple scattering theory~\cite{Drescher:2000ha}. 
EPOS can be subdivided into three main components, (i) the core-corona division, (ii) the hydrodynamical evolution, and (iii) the hadronic cascades.\
 
 (i) The separation of the flux tubes fragmentation into \textit{core} and \textit{corona} (hadronize as a hadron jet) is based on the probability to escape from the bulk matter which will depends on the fragment transverse momentum and the local string density.

(ii)  The hydrodynamical evolution based on the \textit{vHLLE}, viscous HLLE-based algorithm, 3D+1viscous hydrodynamics  employ a realistic Equation of State constrained with Lattice QCD data \cite{Allton:2002zi}.
%

(iii) The hadronic cascade, \textit{hadronic afterburner}, is based on the UrQMD model~\cite{Bleicher:1999xi,Bass:1998ca}, which has been broadly employed to investigate ultra-relativistic heavy-ion collisions~\cite{Bass:1998ca,Bleicher:1999xi,Petersen:2008dd}.
 UrQMD was designed to investigate hadron-hadron, hadron-nucleus, and  heavy-ion collisions from $E_{\rm Lab} = 100$ A$\cdot$MeV to $\sqrt{s_\text{NN}} =$~200 GeV.
Therefore, it includes a collision term that accounts for the interactions of more than 50 (40) baryon (meson) species as well as their anti-particles. The URQMD model describes the hadron-hadron interactions as well as the system evolution based on covariant propagation of all hadrons in the model with resonance decay,  stochastic binary scattering, and color string formation. 
\end{itemize}
The results reported in this work, were obtained for minimum bias Au+Au collisions at $\sqrt{s_{\rm NN}}=$ 200 GeV.  
A total of approximately 4.0, 5.0, 3.0, and 0.3~M events of Au+Au collisions were generated with AMPT Set-1, Set-2, Set-3, and EPOS, respectively.
%

The $\rho(v^{2}_{n},[p_{T}])$ correlator is derived from covariances and variances (cf. Eq.~\ref{eq:1}) which involve both two- and multi-particle correlations that could also be influenced by non-flow effects due to resonance decays, Bose-Einstein correlations, and the fragments of individual jets~\cite{Jia:2013tja}.  
Since  non-flow contributions mostly involve particles emitted within a localized region in pseudorapidity, $\mathrm{\eta}$, they can be mitigated via the sub-event cumulant methods~\cite{Jia:2017hbm,Huo:2017nms,Zhang:2018lls,Magdy:2020bhd}.
 A major mitigating feature of these methods, is the correlation of particles from two or more sub-events which are separated in $\mathrm{\eta}$.
The efficacy of these  methods to reduce non-flow effects have been quantified for many different two- and multi-particle correlators~\cite{Jia:2017hbm,Huo:2017nms,Magdy:2020bhd}. 
It is noteworthy that these methods were used in the initial $\rho(v^{n}_{n},[p_{T}])$ measurements by the ATLAS Collaboration~\cite{Aad:2019fgl}.

In the current work, the two-subevents method is used to construct the $v_{2}^{2}$ variance. Thus, we use two separate $\eta$ selections specified as $-1.0< \eta_{A}<-0.35$  and $0.35<\eta_{C}<1.0$ to determine the $v_{2}^{2}$ variance as:
\begin{eqnarray}\label{eq:2-1}
    {\rm Var}(v_{2}^{2}) &=& v_{2}\{2\}^{4} - v_{2}\{4\}^{4} \nonumber,\\
     &=& C^{2}_{2}\{2\} - C_{2}\{4\},
\end{eqnarray}
where $v_{2}\{2\}$ and $v_{2}\{4\}$ are the flow coefficients obtained from the two- and four-particle correlations respectively, with the sub-event method~\cite{Jia:2017hbm} with particles in the region $\eta_{A}$ and $\eta_{C}$;
\begin{eqnarray}\label{eq:2-2}
C_{2}\{2\} &=&  \langle \langle 2\rangle\rangle|_{A,C}   =  \langle  \langle e^{\textit{i}~ 2 (\varphi^{A}_{1} -  \varphi^{C}_{2} )} \rangle \rangle,\\
v_{2}\{2\}     &=&  \sqrt{C_{n}\{2\}}
\end{eqnarray}
where $\phi_{A(C)}$ is the azimuthal angle of particles in the region $A$ ($C$).
\begin{eqnarray}\label{eq:2-3}
C_{2}\{4\}    &=&  \langle \langle 4\rangle\rangle|_{A,C} - 2 \langle \langle 2\rangle\rangle^{2}|_{A,C},\\
v_{2}\{4\}     &=& \sqrt[4]{-C_{2}\{4\} } 
\end{eqnarray}
where,
\begin{eqnarray}\label{eq:2-4}
\langle \langle 4\rangle\rangle|_{A,C} &=&  \langle  \langle e^{\textit{i}~ 2 (\varphi^{A}_{1} + \varphi^{A}_{2} -  \varphi^{C}_{3} -  \varphi^{C}_{4})} \rangle \rangle.
\end{eqnarray}

The variance of the dynamical $p_{T}$ fluctuations~\cite{Abelev:2014ckr}, $c_k \sim {\rm Var}([p_T])$, defined in region $|\eta_{B}|<0.35$, can be given as:
\small{
\begin{eqnarray}\label{eq:2-5}
    c_k  = \left\langle  \frac{1}{N_{\rm pair}} \sum_{B}\sum_{B^{\prime}\neq B} (p_{T,B} - \langle [p_T] \rangle )  (p_{T,B^{\prime}} - \langle [p_T] \rangle) \right\rangle,  \nonumber \\
\end{eqnarray}
}
where  $\langle  \rangle$ is an average over all events. The event mean $p_T$, $[p_T]$,  is given as,
\begin{eqnarray}\label{eq:2-6}
     [p_T]  =  \sum^{M_{B}}_{i=1} p_{T,i}  /  M_{B},
\end{eqnarray}
where $M_{B}$ is the event multiplicity in  sub-event $B$.
The covariance between $v_{2}^{2}$ and $[p_T]$, ${\rm cov}(v_{2}^{2},[p_T])$,  is defined using the three-subevents method~\cite{Aad:2019fgl,Zhang:2021phk} as,
\small{
\begin{eqnarray}\label{eq:2-7}
{\rm cov}(v_{2}^{2},[p_T]) &=&  {\rm Re} \left( \left< \sum_{A,C} e^{i2(\phi_{A} - \phi_{C})} \left( [p_T] - \langle [p_T] \rangle \right)_{B} \right> \right). \nonumber \\
\end{eqnarray}
}
The  resulting $\rho(v^{2}_{2},[p_{T}])$ correlator, obtained via Eqs.~\ref{eq:2-1}, \ref{eq:2-5} and \ref{eq:2-7};
\begin{eqnarray}\label{eq:2-8}
    \rho(v^{2}_{2},[p_{T}]) = \frac{{\rm cov}(v_{2}^{2},[p_T])}{\sqrt{{\rm Var}(v_{2}^{2})} \sqrt{c_k}},
\end{eqnarray}
 is similar to that used in prior studies~\cite{Giacalone:2020byk,Lim:2021auv,Bozek:2016yoj,Bozek:2020drh,Schenke:2020uqq,Giacalone:2020dln,ATLAS:2021kty}.

\begin{figure}[t] 
\includegraphics[width=1.0 \linewidth, angle=-0,keepaspectratio=true,clip=true]{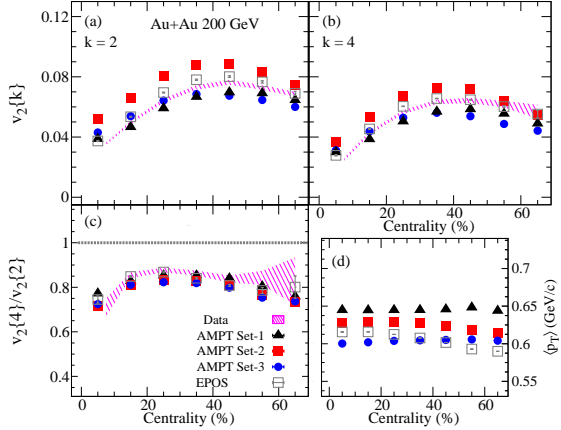}
\vskip -0.4cm
\caption{Centrality dependence of $v_{2}\lbrace 2\rbrace$ (a), $v_{2}\lbrace 4\rbrace$ (b), $v_{2}\lbrace 4\rbrace / v_{2}\lbrace 2\rbrace$ (c)  and $\left< p_T \right>$ computed with the AMPT and EPOS models for Au+Au collisions at 200~GeV. The hatched bands represent the experimental data reported by the STAR collaboration in Ref~\cite{Adams:2004bi}. 
  }\label{fig:1}
\end{figure}

The $\rho(v^{2}_{n},[p_{T}])$ correlator is derived from the correlations and fluctuations of $v_n$ and $p_T$. Therefore, it is instructive to investigate the dependence of these variables on the models and the corresponding parameters tabulated in Table~\ref{tab:1}. Fig. ~\ref{fig:1} shows a comparison of the centrality dependence of  $v_{2}\lbrace 2\rbrace$ (a), $v_{2}\lbrace 4\rbrace$ (b) the ratios $v_{2}\lbrace 4\rbrace/v_{2}\lbrace 2\rbrace$ (c), and $\left< p_T \right>$ (d) for the AMPT and EPOS models. Panels (a) and (b) indicate that the AMPT results are sensitive to both the viscosity and whether or not string melting is turned on. They also indicate similar qualitative patterns between both models and the data reported by the STAR collaboration~\cite{Adams:2004bi} (hatched bands) over the range of the model parameters summarized in Table~\ref{tab:1}. Note that the experimental results are only shown for published data. 
{\color{black} An extraction of  the $\langle p_{\rm T} \rangle$ from the experimental charged hadrons spectra reported in Ref~\cite{STAR:2003fka}, indicates values which are  closer to those obtained with the EPOS model.}

The ratios $v_{2}\{4\}/ v_{2}\{2\}$, shown in panel (c), serve as a figure of merit for the strength of the elliptic flow fluctuations; $v_{2}\{4\}/ v_{2}\{2\} = r$  correspond to minimal flow fluctuations for $r \sim 1.0$ and sizable flow fluctuations for decreasing values of $r < 1$. The computed ratios, which are similar to the experimental constructed ones, are to within$\sim $5\% insensitive to the model choice and the parameter sets tabulated in Tab.~\ref{tab:1}, suggesting that the flow fluctuations are eccentricity-driven and are roughly a constant fraction of $v_{2}\{2\}$. The $\left< p_T \right>$ values shown in panel (d) further indicate minor sensitivity to the models and the associated model parameters.
%

%
\begin{figure}[hbt]
    \includegraphics[width=1.0 \linewidth, angle=-0,keepaspectratio=true,clip=true]{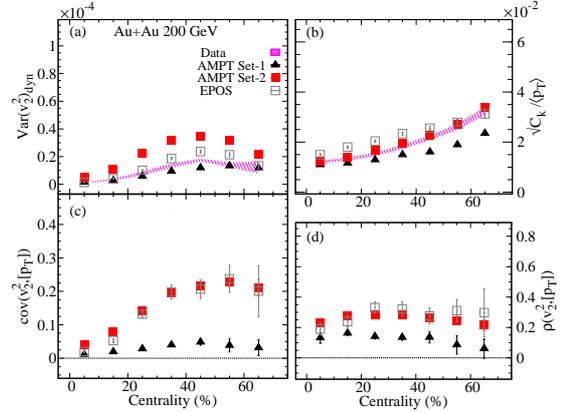}
      \vskip -0.4cm
    \caption{    
  Comparison of the centrality dependence of ${\rm Var}(v_{n}^{2})_{dyn}$ (a), $\sqrt{c_{k}}/\langle p_{T}\rangle$ (b), ${\rm cov}(v_{n}^{2},[p_T])$ (c) and $\rho(v^{2}_{n},[p_{T}])$ (d) obtained from the AMPT and EPOS models for Au+Au collisions at 200~GeV. The hatched bands represent the experimental data reported by the STAR collaboration in Refs~\cite{Adams:2004bi,STAR:2019dow}. \label{fig:2}
  }
\end{figure}

Figure~\ref{fig:2} compare the centrality dependence of the values for ${\rm Var}(v_{2}^{2})$ (a),  $\sqrt{c_{k}}/\langle p_{T}\rangle$ (b), ${\rm cov}(v_{2}^{2},[p_T])$ (c) and $\rho(v^{2}_{2},[p_{T}])$ (d), computed for tracks with $0.2<p_{T}<2.0$~\GeVc in Au+Au collisions simulated with the AMPT and EPOS models. 
 The hatched bands represent ${\rm Var}(v_{2}^{2})$ (a) and $\sqrt{c_{k}}/\langle p_{T}\rangle$ (b) values evaluated from the experimental data reported by the STAR collaboration~\cite{Adams:2004bi,STAR:2019dow}. They show good qualitative agreement with the theoretical values for ${\rm Var}(v_{2}^{2})$, and $\sqrt{c_{k}}/\langle p_{T}\rangle$. 
The comparisons show good overall agreement between the results for EPOS and those for AMPT with parameter Set-2.
By contrast, panels (a) - (d) indicate sizable differences between the AMPT results for parameter Set-1 [string melting off] and Set-2 [string melting on]. These differences suggest that data-model comparisons of ${\rm Var}(v_{n}^{2})_{dyn}$, $\sqrt{c_{k}}/\langle p_{T}\rangle$, ${\rm cov}(v_{n}^{2},[p_T])$ and $\rho(v^{2}_{n},[p_{T}])$ could serve as an important constraint for charting the respective roles of the partonic and hadronic phases in these collisions. 
\begin{figure}[hbt]
  \vskip -0.2cm
    \centering
\includegraphics[width=1.0 \linewidth, angle=-0,keepaspectratio=true,clip=true]{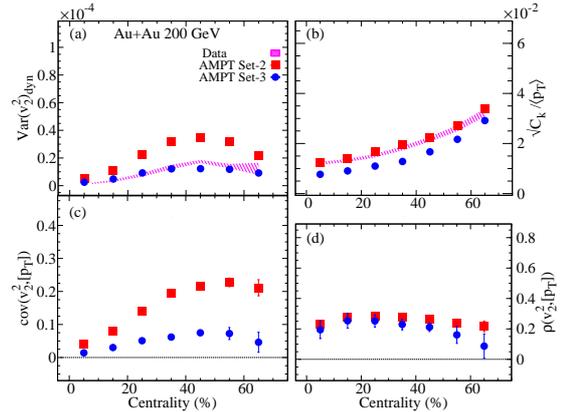} 
\vskip -0.4cm
     \caption{
		Comparison of the centrality dependence of the values for ${\rm Var}(v_{n}^{2})_{dyn}$ (a), $\sqrt{c_{k}}/\langle p_{T}\rangle$ (b), ${\rm cov}(v_{n}^{2},[p_T])$ (c) and $\rho(v^{2}_{n},[p_{T}])$ (d), computed for Au+Au collisions at 200~GeV with the AMPT model.  Results are compared for model parameter Set-2 ($\eta/s$ = 0.1) and Set-3 ($\eta/s$ = 0.3) as indicated. The hatched bands represent the experimental data reported by the STAR collaboration in Refs~\cite{Adams:2004bi,STAR:2019dow}.
               }
    \label{fig:6} 
  \vskip -0.2cm
\end{figure}

\begin{figure}[ht] 
    \includegraphics[width=1.0 \linewidth, angle=-0,keepaspectratio=true,clip=true]{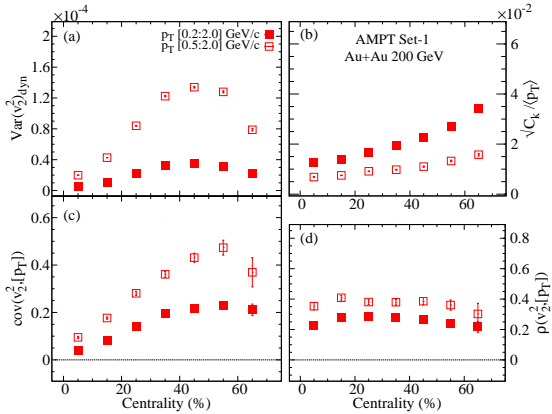}
    \vskip -0.4cm
    \caption{
Comparison of the centrality-dependent values for ${\rm Var}(v_{n}^{2})_{dyn}$ (a), $\sqrt{c_{k}}/\langle p_{T}\rangle$ (b), ${\rm cov}(v_{n}^{2},[p_T])$ (c) and $\rho(v^{2}_{n},[p_{T}])$ (d) computed for the selections $0.2<p_{T}<2.0$~\GeVc and  $0.5<p_{T}<2.0$~\GeVc for Au+Au collisions at 200~GeV with the AMPT model .
  } \label{fig:3}
    \vskip -0.0cm
\end{figure}

Figure~\ref{fig:6} shows the influence of $\eta/s$ on the values for ${\rm Var}(v_{2}^{2})$ (a), $\sqrt{c_{k}}/\langle p_{T}\rangle$ (b), ${\rm cov}(v_{2}^{2},[p_T])$ (c) and $\rho(v^{2}_{2},[p_{T}])$ (d). Here, results from the AMPT model are shown for model parameter Set-2 ($\eta/s$ = 0.1) and Set-3 ($\eta/s$ = 0.3) as indicated. 
The hatched bands represent the values for ${\rm Var}(v_{2}^{2})$ (a) and $\sqrt{c_{k}}/\langle p_{T}\rangle$ (b) constructed from the experimental data reported by the STAR collaboration~\cite{Adams:2004bi,STAR:2019dow}.
Figs.~\ref{fig:6} (a) - (c) show that an increase in the magnitude of \etas reduces the values for ${\rm Var}(v_{2}^{2})$, $\sqrt{c_{k}}/\langle p_{T}\rangle$, and ${\rm cov}(v_{2}^{2},[p_T])$ for all centrality selections. By contrast,  the $\rho(v^{2}_{2},[p_{T}])$ values show little, if any, change, suggesting that an apparent cancellation of the effects of viscous attenuation renders the $\rho(v^{2}_{2},[p_{T}])$ correlator insensitive to $\eta/s$.

Differential flow measurements $v_{n}(p_T)$ indicate that $v_n$ is strongly correlated with $p_T$. Consequently, it is instructive to investigate the influence of a $p_T$ cut on the $\rho(v^{2}_{2},[p_{T}])$ correlator. Fig.~\ref{fig:3} compare the extracted values for ${\rm Var}(v_{2}^{2})_{dyn}$ (a), $c_k$ (b), ${\rm cov}(v_{2}^{2},[p_T])$ (c) and $\rho(v^{2}_{2},[p_{T}])$ (d) computed for the selections $0.2<p_{T}<2.0$~\GeVc and $0.5<p_{T}<2.0$~\GeVc in Au+Au collisions ($\sqrt{s_{\rm NN}}$= 200 GeV) with the AMPT model.
Figs.~\ref{fig:3} (a), (c) and (d) indicate a sizable increase in the centrality-dependent values for ${\rm Var}(v_{2}^{2})_{dyn}$, ${\rm cov}(v_{2}^{2},[p_T])$ and $\rho(v^{2}_{2},[p_{T}])$ with $\left< p_T \right>$~\cite{Aad:2019fgl}. Contrastingly, the centrality-dependent values for $c_k$ indicate a decrease with $\left< p_T \right>$. 

The sensitivity of the $\rho(v^{2}_{n},[p_{T}])$ correlator to the shape and size of the collision system was investigated using the  Event Shape Engineering (ESE) technique~\cite{Adler:2002pu}. This technique leverages the observation that selections on the magnitude of the event-by-event fluctuations of the  $v_{n}$ coefficients, for a fixed centrality, serve to influence the shape of the collision system~\cite{Abelev:2012di}.  
\begin{figure}[hbt]
  \vskip -0.3cm
  \center{
\includegraphics[width=0.7 \linewidth,angle=-0,keepaspectratio=true,clip=true]{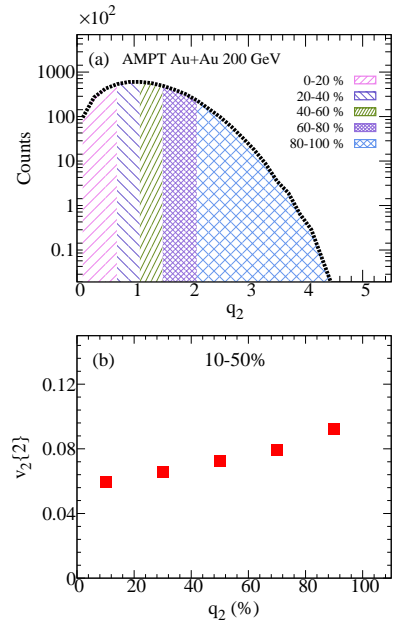}
  \vskip -0.5cm
 \caption{
 The $q_{2}$ distribution (a) for $10-50\%$ central Au+Au collisions at $\sqrt{s_{NN}}$ = 200~GeV, obtained with the sub-event cut $\mathrm{1.5<\eta < 2.5}$ with the AMPT model; $ v_{2}\lbrace2\rbrace$  vs. $q_{2} (\%)$ (b) for the $q_2$ selections indicated in panel (a).
 } \label{fig:4} 
   \vskip -0.2cm
   }
\end{figure}
\begin{figure}[hbt]
  \vskip -0.3cm
\includegraphics[width=1.0 \linewidth,angle=-0,keepaspectratio=true,clip=true]{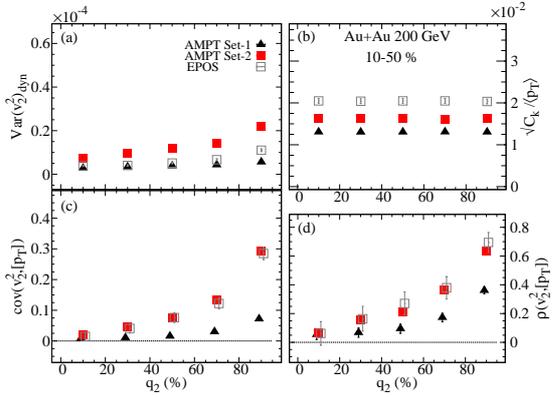}
  \vskip -0.5cm
 \caption{
Comparison of the  $q_{2}$-dependent of ${\rm Var}(v_{n}^{2})_{dyn}$ (a), $\sqrt{c_{k}}/\langle p_{T}\rangle$ (b), ${\rm cov}(v_{n}^{2},[p_T])$ (c) and $\rho(v^{2}_{n},[p_{T}])$ (d) computed for  $10-50\%$ central Au+Au collisions ($\sqrt{s_{NN}}$ = 200~GeV) obtained with the AMPT and EPOS models. 
 } \label{fig:5} 
   \vskip -0.2cm
\end{figure}
%

The event-shape selections were performed via a fractional cut on the distribution of the magnitude of the reduced second-order flow vector, $q_2$,~\cite{Schukraft:2012ah,Adler:2002pu};
%
\begin{eqnarray}
Q_{2, x} = \sum_{i} \cos(2 \varphi_{i}),
Q_{2, y} = \sum_{i} \sin(2 \varphi_{i}),
\end{eqnarray}
\begin{eqnarray}
q_{2}    &=& \frac{|{Q}_{2}|}{\sqrt{M}}, ~|Q_{2}|  = \sqrt{Q_{2, x}^2 + Q_{2, y}^2}
\end{eqnarray}
where $Q_{2}$ is the magnitude of the second-order harmonic flow vector calculated within  the sub-event $\mathrm{1.5< \eta < 2.5}$, and $M$ is the charged hadron multiplicity for this sub-event. The sub-event $|\eta| < 1.0$  was used to evaluate $\rho(v^{2}_{n},[p_{T}])$ to ensure the separation between the sub-event used to evaluate $q_{2}$ and $\rho(v^{2}_{n},[p_{T}])$. 
Note that there are two caveats to the ESE method. First, the $q_{2}$ selective power depends on the magnitude of $v_{2}$ and the event multiplicity, so the benefit of the technique is handicapped by weak flow values and small event multiplicities~\citep{Bzdak:2019pkr}. 
Second, the non-flow effects, such as resonance decays, jets, etc.~\citep{Voloshin:2008dg}, could bias the $q_{2}$ selections. The latter can be minimized via a $\Delta\eta$ separation between the sub-events used for the $q_{2}$ selections and $\rho(v^{2}_{n},[p_{T}])$ evaluations.
Fig.~\ref{fig:4} (a) shows a representative $q_{2}$ distribution for 10-50\% central Au+Au collisions at 200~GeV. The $v_{2}\{2\}$ values which result from the $q_2\%$ selections indicated in panel (a), are shown in Fig.~\ref{fig:4} (b). They indicate an essentially linear increase of $v_{2}\{2\}$ with $q_2\%$.

Figure~\ref{fig:5} compare the $q_2\%$ dependence of the values for ${\rm Var}(v_{2}^{2})$ (a), $\sqrt{c_{k}}/\langle p_{T}\rangle$ (b), ${\rm cov}(v_{2}^{2},[p_T])$ (c) and $\rho(v^{2}_{2},[p_{T}])$ (d), computed for tracks with $0.2<p_{T}<2.0$~\GeVc in Au+Au collisions simulated with the AMPT and EPOS models. A striking feature of these results is the $q_2\%$-independence of $c_k$ and the essentially quadratic dependence of ${\rm cov}(v_{2}^{2},[p_T])$  and $\rho(v^{2}_{2},[p_{T}])$ on $q_2\%$. These dependencies suggests that data-model comparisons of the $\rho(v^{2}_{2},[p_{T}])$ correlators extracted in shape-engineered events, could serve as a sensitive constraint for the initial-state eccentricity, and give important insight on the deformation of colliding systems.

In summary, we have presented extensive model studies to evaluate the model dependence, as well as the response and sensitivity of the $\rho(v^{2}_{2},[p_{T}])$ correlator to collision-system size and shape and the \etas of the matter produced in the collisions.
We find that $\rho(v^{2}_{2},[p_{T}])$ is sensitive to the event shape selections of the collision system, but insensitive to sizable changes in the \etas of the medium produced in the collisions.  
Initial comparisons of the calculated and experimental ${\rm Var}(v_{2}^{2})$, and $\sqrt{c_{k}}/\langle p_{T}\rangle$ values, also indicate good qualitative agreement.
These findings strongly suggest that precise differential measurements of $\rho(v^{2}_{2},[p_{T}])$ as a function of system-size, shape, deformation and beam-energy could provide more stringent constraints to discern between initial-state models and hence, more reliable extractions of $\eta/s$.
%
\section*{Acknowledgments}
The authors thank Marysia Stefaniak, Emily Racow, Benjamin Schweid, and Giuliano Giacalone  for  useful discussions. 
This research is supported by the US Department of Energy, Office of Nuclear Physics (DOE NP),  under contracts DE-FG02-94ER40865 (NM) and DE-FG02-87ER40331.A008 (RL).

%


\bibliographystyle{elsarticle-num}
\bibliography{ref} 
\end{document}